\documentclass[10pt,twocolumn]{article}

\usepackage{graphicx}
\usepackage{amssymb}
\usepackage{mathtools, cuted}
\usepackage{url}
\usepackage[super,comma]{natbib}
\usepackage{fullpage}

\newcommand*{\Scale}[2][4]{\scalebox{#1}{$#2$}}%

\title{Distinguishing deformation mechanisms in elastocapillary experiments}

\author{
Shih-Yuan Chen,\textit{$^{a}$} Aaron Bardall,\textit{$^{b}$} Michael Shearer,\textit{$^{b}$} and Karen E. Daniels\textit{$^{a}$}\\
  {\textit{$^{a}$} \textit{Department of Physics, North Carolina State University}}\\
  {\textit{$^{b}$} \textit{Department of Mathematics, North Carolina State University}}\\
  }

\begin{document}
\twocolumn[
  \begin{@twocolumnfalse}
    \maketitle
    \begin{abstract}
      Soft materials are known to deform due to a variety of mechanisms, including capillarity, buoyancy, and swelling. In this paper, we present experiments on polyvinylsiloxane gel threads partially-immersed in three liquids with different solubility, wettability, and swellability. Our results demonstrate that deformations due to capillarity, buoyancy, and swelling can be of similar magnitude as such threads come to static equilibrium. To account for all three effects being present in a single system, we derive a model capable of explaining the observed data and use it to determine the force law at the three-phase contact line. The results show that the measured forces are consistent with the expected Young-Dupr\'{e} equation, and do not require the inclusion of a tangential contact line force.
    \end{abstract}

\vspace*{20pt}
  \end{@twocolumnfalse}
]


\section{Introduction}

Elastocapillarity is a study of how surface tension  
deforms soft materials such as polymers and biological tissues,\cite{review:Liu, review:Roman, review:Bico, review:Andreotti, cite:Andreotti_shuttle, review:Style} giving rise to such phenomena as  
film wrinkles,\cite{cite:Huang_2007, cite:Benny_2018}
the clumping of wet hairs, \cite{cite:Bico_2004}
capillary origami, \cite{cite:Py_2007, cite:Legrain_2014}
substrate deformation due to partial wetting,\cite{review:Andreotti, review:Style, cite:Jerison_ft, cite:Style_angle, cite:Bostwick, cite:Aaron_hankel}
differences in advancing and receding contact angles,\cite{cite:Kajiya, cite:Zhao_2018}
the Shuttleworth effect,\cite{cite:Xu_shuttle, cite:Liang_2018_Langmuir}
and more. An improved, quantitative understanding of elastocapillary effects is crucial to explaining various industrial and biological phenomena, including 
soft stamp deformation,\cite{cite:Hui}
high-aspect-ratio polymer pillars,\cite{cite:Chandra_2008}
bending flexible legs of water striders,\cite{cite:Park_2008} and
passive droplet motion\cite{cite:Bueno_2017_EML} including  durotaxis\cite{cite:Style_2013_PNAS}.
A complete description of elastocapillary deformation will include all external and internal forces on the soft material. These can simultaneously include
adhesion,\cite{review:Style}
hydrostatic pressure,\cite{cite:Marchand}
electric forces,\cite{cite:Pineirua}
surface tension changes due to uncrossliked oligomers from silicone,\cite{cite:Fargette_oil,cite:Fargette_oil_2, thesis:Matthew}
and  swelling\cite{cite:Whitesides, cite:Wang, cite:Honda, cite:Jan}.
These various mechanisms act within the bulk, or only at an interface.
In any elastocapillary experiment, it is necessary to quantify both the relative magnitude of these forces, and the timescales over which they act, in order to develop a valid, predictive model.

In this paper, we examine how hydrostatic forces and swelling dynamics, commonly neglected in many elastocapillary experiments, can significantly influence the observation of contact line forces. Buoyancy is present for any material immersed in a liquid, and swelling often occurs as the liquid is absorbed by the gel, causing it to increase its volume\cite{review:Dervaux_2012}.  To quantitatively resolve these two effects, we perform experiments on polyvinylsiloxane (PVS) gel threads partially immersed in three representative liquids. These are chosen to be ethanol (polar and amphiphilic), glycerol (polar and hydrophilic), and Fluorinert FC-40 (nonpolar and hydrophobic). These liquids cover a wide range of physical and chemical traits and are typical choices in many previously-performed elastocapillary experiments\cite{cite:Marchand, cite:Style_angle, cite:Schulmann_shuttle}.
We observe that the equilibration of internal stresses takes place over several hours, regardless of the choice of fluid. In contrast, deformation due to swelling is a fluid-dependent effect: for ethanol we observe that it dominates over buoyancy-induced deformations, while for glycerol or Fluorinert swelling effects are small. This observation is consistent with the work of \citet{cite:Whitesides}, in which it was observed that the swelling ratio of ethanol was an order of magnitude larger than for glycerol or Fluorinert.

To quantify and understand these deformations, we develop a model that includes all three effects -- capillarity, buoyancy, and swelling. In what follows, we use the term {\it surface tension} to describe the capillary force at the liquid-air interface, and {\it surface stress} to describe the capillary force at the solid-air and  solid-liquid interfaces. Our model successfully reproduces the observed deformations with only surface stress and a small swelling ratio as free parameters. Using this model, we are able to address a current controversy surrounding the modeling of elastocapillary wetting: the presence or absence of a tangential component of the contact line force\cite{review:Andreotti, cite:Jerison_ft, cite:Style_2012, cite:Bostwick, cite:Weijs_tangen, cite:Marchand, cite:Marchand_2011, cite:Neukirch_2014, cite:Park_visual}. 
In previous experiments,\cite{cite:Marchand} it was reported that a partially-immersed thread experiences internal deformations consistent with the presence of a component of the contact line force which is tangential to the substrate surface:
\begin{equation}
 F_{\mathrm{\parallel}}=\gamma(1+\cos\theta),
\end{equation}
where $\gamma$ is the surface tension of the liquid, and $\theta$ is the Young's angle. The presence of this tangential component, hypothesized to arise from the liquid molecules attracting polymers at the surface of the thread via van der Waals forces, would mean that the contact line for a soft solid follows neither the Young-Dupr\'e equation nor Neumann's equation. Importantly, the experiments of \citet{cite:Marchand} were done using ethanol as the immersion fluid, while  studies using water, glycerol, or Fluorinert droplets\cite{ cite:Style_angle, cite:Bostwick, cite:Style_2012, cite:Park_visual}  deposited on a gel substrate found no necessity for introducing a tangential component for the contact line force in order to explain the observed deformations.
In previous experiments on a thread immersed in ethanol,\cite{cite:Marchand} swelling and buoyancy effects were incompletely included in the explanatory model. In this work, we take into account both of these forces, together with the viscous equilibration of beads within the thread, and thereby observe that the contact line force law is in agreement with with the Young-Dupr\'e equation, consistent with previous work on droplets\cite{cite:Style_angle,cite:Bostwick,cite:Style_2012,cite:Park_visual}. 

\section{Methods \label{sec:method}}

\subsection{Apparatus}

\begin{figure}
\begin{center}
\includegraphics[width=0.45\textwidth]{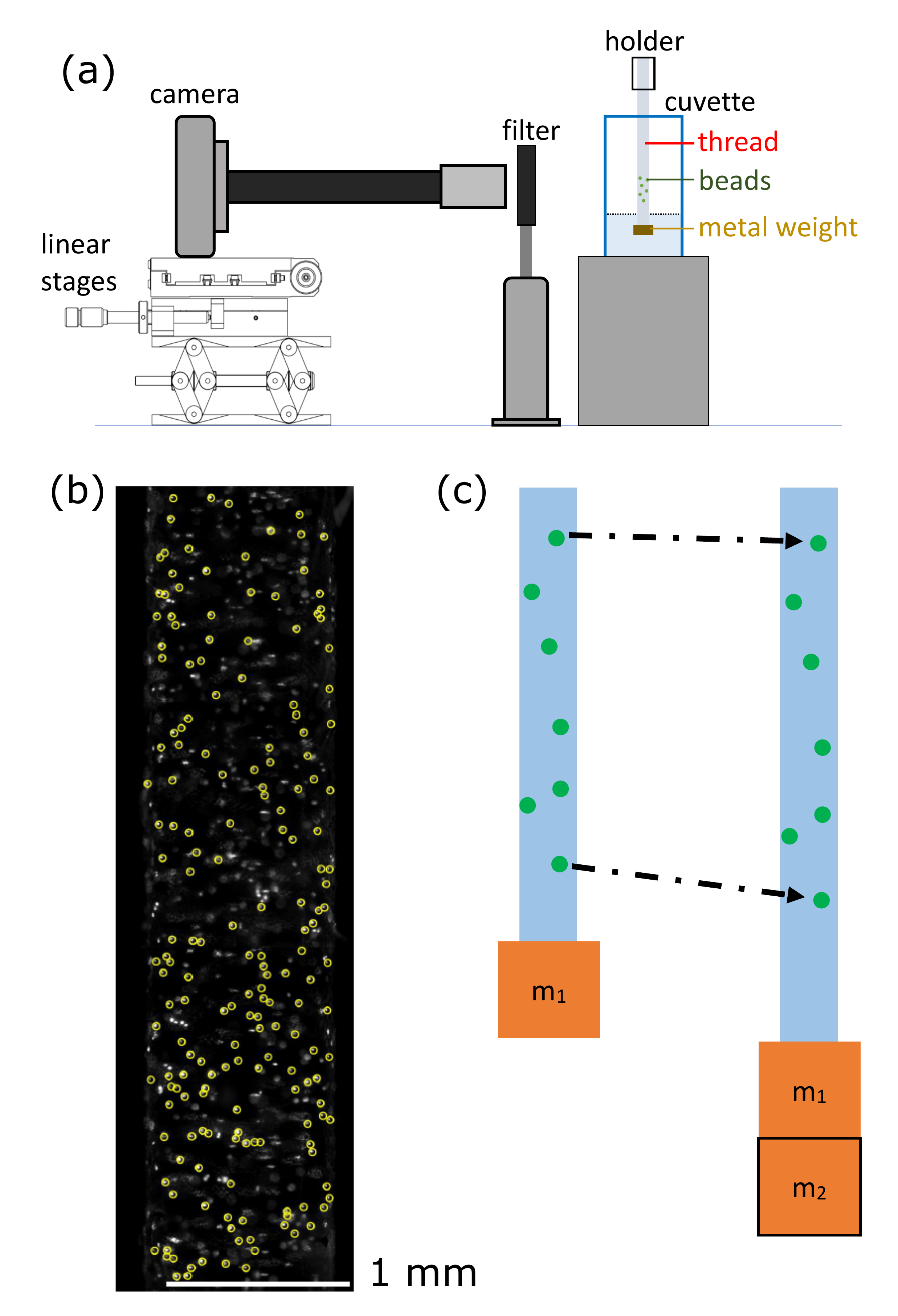}
\end{center}
\caption{(a) Schematic of the experiment setup, showing the placement of the gel thread within the cuvette. (b) A sample image of fluorescent beads located within the gel thread. Each yellow circle marks a bead used to track displacements. (c) Schematic of the response of the fluorescent beads to a change in stress, in this case due to a change in the mass of the metal weights from $m_1$ to $m_1 + m_2$. Similar displacement measurements are used in all experiments, for measuring the response due to swelling, buoyancy, and capillary force.}
\label{fig:setup}
\end{figure}

As was done in the prior experiments by \citet{cite:Marchand}, we measure the internal deformation of a polyvinylsiloxane (PVS) gel thread immersed in a fluid; the casting process for the threads is described in \S\ref{subsec:gel}. The thread is lowered into a glass cuvette  via a linear stage (Thorlabs), and then one of three fluids (ethanol, glycerol, Fluorinert) is added; a schematic is shown in Fig.~\ref{fig:setup}a. The size of the cuvette ($2$~cm $\times$ $5$~cm $\times$ $5$~cm) is chosen to be large enough that the liquid surface can be assumed to be  flat at the center.

Internal deformations are observed via fluorescent beads ($525$~nm absorption, $565$~nm emission) cast into the gel, serving as position markers. A green laser (Laserglow, $532$~nm) illuminates the beads, and the emitted light passes through a notch filter (Edmund, $568$~nm with $2$~nm FWHM) before being recorded on a low-noise digital CCD (Andor Luca R 604, $1004 \times 1002$ pixels). In combination with the lens (10$\times$ Mitutoyo long working distance objective), the resolution is $1.6 \ \mu$m/pixel, and the depth of field is around 1 bead diameter. 
Since the thread is longer than the field of view ($1.6$~mm) of our camera, we take pictures of the thread by moving the camera along its length using a 3-axis linear stage (Thorlabs) with both coarse and fine control. 
We locate the position of each bead using a computer code built around the MATLAB \texttt{regionprops} algorithm, and then we stitch each series of images into a single image by maximizing the cross correlation between two adjacent images. A sample result is shown in Fig.~\ref{fig:setup}b.

\subsection{Gel thread preparation and characterization 
\label{subsec:gel}}

Each  gel thread is  cast inside a glass capillary tube (Wheaton, $5-25 \ \mu$L) from 2-part PVS (Ecoflex 00-10). To prevent sticking between the gel and the capillary tube, we silanize the inner and outer surfaces of the tube  using a solution of $5\%$ (tridecafluoro-1,1,2,2-tetrahydrooctyl)trichlorosilane (Gelest) in $95\%$ 2-Propanol (Sigma-Aldrich) and dry it in an oven for 30 minutes at $150^{\circ}$C.  Fluorescent polystyrene beads (envy green, Bangs Laboratories, $10 \ \mu$m in diameter) are mixed into one component of the gel mixture at a concentration of $10 \ \mu$L$/$g (volume of bead solution per weight of liquid). Immediately prior to casting, we use a petri dish to mix the first and second components (1:1 weight ratio), and degas the mixture in a vacuum chamber  until there are no visible bubbles (typically at least 3 minutes). We draw the gel-bead mixture into the capillary tube using a syringe pump. The threads are cured at room temperature for 24 hours, after which we break the capillary tube and pull it away to reveal the cured thread. The resulting thread has a length $L= 2.5$~cm, and radius $R= 475 \, \mu$m. The density of the gel is $\rho_s= 1040$~kg/m$^3$. This one thread is used for all of our experimental trials.

To ensure that the thread hangs straight in all experiments, we glue a small metal weight ($0.38 \pm 0.01$ g, shown as $m_1$ in Fig.~\ref{fig:setup}c) to the lower end of the thread using a small amount of PVS gel. To measure the Young's modulus of the thread, we temporarily attach an additional metal weight  $m_2= 0.09$, $0.16$, $0.21$, or $0.29$~g to the thread, as shown in Fig.~\ref{fig:setup}c, further stretching the thread. From the displacement fields made by comparing the positions of beads under pairs of loading conditions, we find a linear relationship and measure the average Young's modulus to be $E = 100 \pm 15$~kPa, with the error bounds given by the standard error across 7 pairwise measurements. All Young's modulus measurements were done in air.

\subsection{Cleaning procedure \label{subsec:clean}}

Typically, a cured elastomer will contain remnants of uncrosslinked polymer oil known as oligomers; when the elastomer is in contact with a liquid, some of these oligomers will  dissolve into the liquid\cite{cite:Whitesides}. These oligomers, when they migrate to the surface of the liquid, have been observed to act as a surfactant and thereby reduce the surface tension of the liquid\cite{cite:Fargette_oil,cite:Fargette_oil_2,thesis:Matthew}.
In order to perform controlled experiments, we follow the cleaning procedure suggested by \citet{cite:Fargette_oil}, modified in duration to account for the use of ethanol in place of toluene. During two sequential 24-hour immersions in ethanol, the oligomers dissolve from the thread into the bath and are flushed away. Finally, we dry the thread in a vacuum chamber for 15 minutes to remove the residual ethanol, followed by air-drying for a week.

\subsection{Liquids \label{subsec:liquid}}

Based on previous elastocapillary experiments,\cite{cite:Style_angle, cite:Marchand, cite:Schulmann_shuttle} we select the three representative liquids (ethanol, glycerol, Fluorinert FC-40), the detailed properties of which are given in Table \ref{tab:liquid}.  All liquids are purchased from Sigma-Aldrich; we measure the  density $\rho_l$ using a pycnometer (Kimble, 10 mL) and the liquid surface tension $\gamma$ using a tensiometer (Surface Tensionmat, Fisher). To measure the Young's angle $\theta$ for PVS, we first cast a thick, flat slab and clean it using the methods given in \S\ref{subsec:clean}. We deposit droplets of diameter $D \gtrsim 3$~mm, so that the ratio $\gamma/ED \ll 1$ is satisfied, image each droplet from the side, and measure $\theta$ using ImageJ. Errors reported are the standard error across 6 measured droplets. 
Finally, to quantify the {\it in-situ} evaporation rate for each liquid, we measure the combined mass of the liquid and cuvette over a period of 3 hours, and record the loss. 

\begin{table*}
	\centering
\begin{tabular}{|l|c|c|c|}
\hline
Liquid                                                                   & 95\% ethanol          & glycerol        & \begin{tabular}[c]{@{}c@{}}Fluorinert\\ FC-40\end{tabular} \\ \hline
Density $\rho_l$                                                         & $0.8 $~g/cm$^3$       & $1.25$~g/cm$^3$ & $1.855$~g/cm$^3$ \\ \hline
\begin{tabular}[c]{@{}l@{}}Surface tension\\$\gamma$\end{tabular} & $22$~mN/m          & $60$~mN/m       & $16$~mN/m \\ \hline
\begin{tabular}[c]{@{}l@{}}Young's angle\\$\theta$ to PVS \end{tabular} & $\approx 45^{\circ}$ & $106^{\circ} \pm 6^{\circ} $ & $ 29^{\circ} \pm 3^{\circ} $ \\ \hline
\begin{tabular}[c]{@{}l@{}}Evaporation\\rate in cuvette\end{tabular}     & $100 \ \mu $m/hr & $<1 \ \mu $m/hr & $10 \ \mu $m/hr \\ \hline
& polar  & polar & nonpolar \\ \hline
& amphiphilic  & hydrophilic & hydrophobic \\ \hline
\end{tabular}
\caption{The three liquids used in the experiments. Values are measured as described in \S\ref{subsec:liquid}, except for the Young's angle $\theta$ of ethanol, which is given by \citet{cite:Marchand}.}
\label{tab:liquid}
\end{table*}

\subsection{Observing procedure \label{subsec:protocol}}

\begin{figure}
\begin{center}
\includegraphics[width=0.45\textwidth]{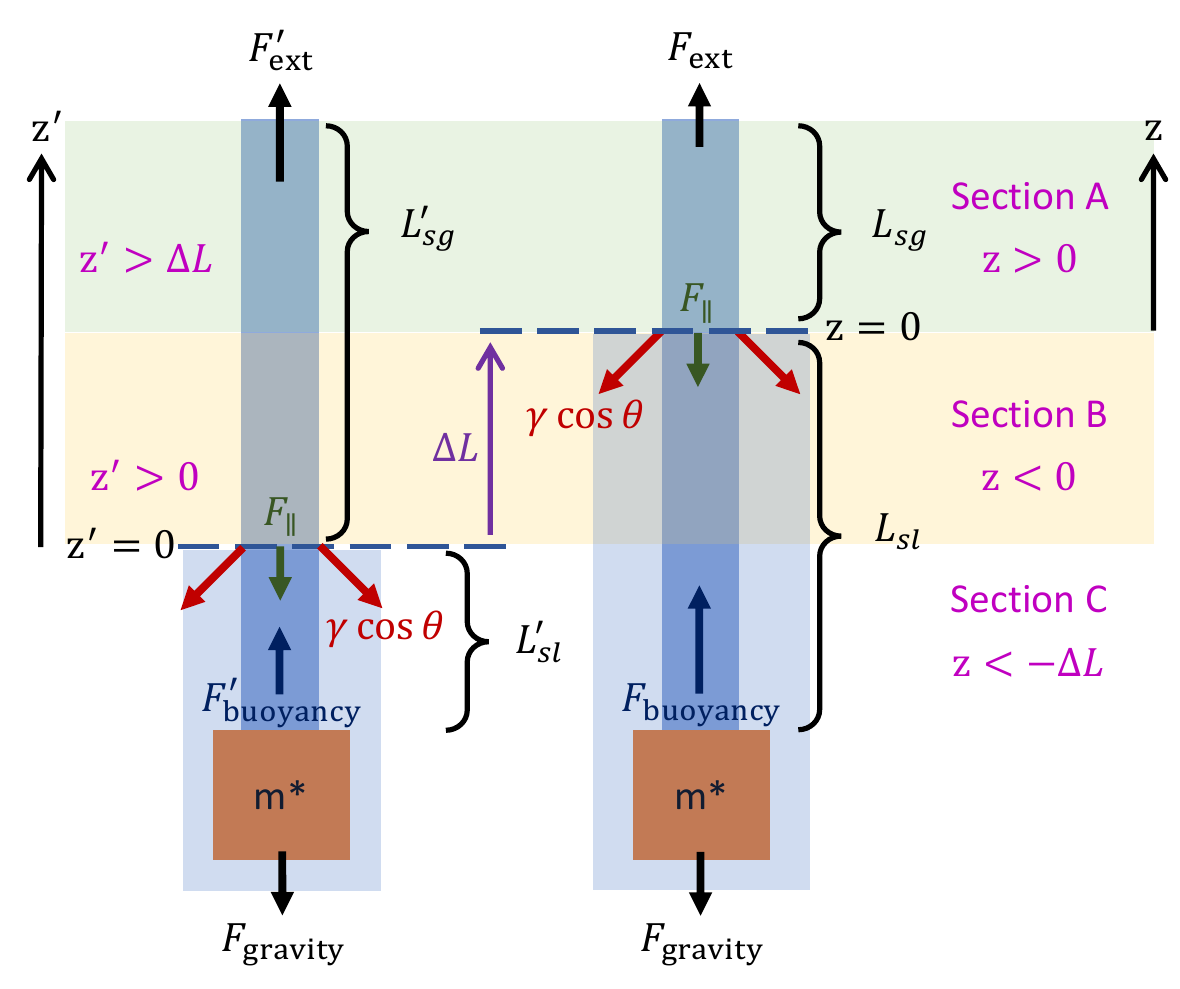}
\end{center}
\caption{Schematic diagram for the observing procedure, with the left side representing the initial state (primed variables, coordinate system $z'$) and the right side representing the final state (un-primed variables, coordinate system $z$). The two coordinate systems are displaced relative to each other by a distance $\Delta L$, the imposed increase in the liquid level between the initial and finals states of the experiment. The horizontal dashed line separates the segments above (length ${L}'_{sg}$) and below (length ${L}'_{sl}$) the liquid surface. The thread has a metal weight of effective mass $m^{*}$ (modified by buoyancy) hanging from its end. Static force equilibrium results from balancing several stresses (denoted $F$, measured in mN/m) acting on the thread: the stress at the attachment point is $F_{\mathrm{ext}}$, which equals the vector sum of the stress $F_{\mathrm{gravity}}$ from the gravitational force,  $F_{\mathrm{buoyancy}}$ from buoyancy, $\gamma \cos \theta$ from surface tension, and $F_{\mathrm{\parallel}}$ from the tangential component of the contact line force.}
\label{fig:exp}
\end{figure}

As summarized schematically in Fig.~\ref{fig:exp}, we start from a state in which the gel thread is hanging, centered, in a cuvette (not shown). We pour in the chosen liquid until the liquid level is higher than the position of the metal weight and then wait for a specified waiting time, $t_w$ (ranging from 0 to 24 hrs). The position of the liquid surface is denoted $z'=0$, and all primed variables are in reference to this initial state. Next, we perform a vertical scan of images for recording the bead positions, and then increase the liquid level by a specified $\Delta L$ to $z=0$  and again wait for time $t_w$. The un-primed variables refer to this final state. We then perform a vertical scan of images for recording bead positions, and use particle tracking \cite{cite:web} to measure the displacements of the beads between the final and initial states.

\section{Modelling \label{sec:theo}}

Our model is based on the differences induced within the gel thread,  between the initial and final configurations shown schematically in Fig.~\ref{fig:exp}. Both the left and right threads are in static equilibrium, with the only external change being the liquid level. We specify a different coordinate system for the two states, with the initial (left) state specified by cylindrical  coordinates $\left(r', \phi', {z}' \right)$, and the final (right) state by  $\left(r, \phi, z \right)$. For an increase in liquid level by a height $\Delta L$,  $r'=r$, $\phi' = \phi$, and ${z}'= z + \Delta L$. Thread segments of length ${L}'_{sg}$ (above the liquid surface) and ${L}'_{sl}$ (below the liquid surface) add up to the total length $L'$. We further assume that $L' = L$ and radius $R' = R$ are  approximately constant for the two states; note that our bead displacements are around $25 \, \mu$m, and the length of the thread is 25 mm (see \S\ref{sec:result}).

Two segments of the thread form the focus of our work: Segment A ($\bf A$bove the liquid in both states) and Segment B ($\bf B$elow the liquid in the final state); we ignore the segment that remains below the liquid surface in both states. Segment A therefore has ${z}'>\Delta L$ in the initial state and $z>0$ in the final state, while Segment B has ${z}'>0$ in the initial state and $z<0$ in the final state. 

To analyze the different deformations present in the initial and final states, we derive the force balance equation that holds a partially immersed thread in place using the principle of virtual work. From this formulation, we compare the strain field in the initial state (${\epsilon}'$) to the strain field the final state ($\epsilon$), via the {\it strain difference}, $\Delta \epsilon= \epsilon-{\epsilon}'$. The integration of the strain difference gives us the displacement field $u$ in the thread. Note that the forces to be considered in calculating $u$ for Segment A differ from those required in Segment B, due to the absence/present of the liquid.

In writing our model, we consider two forces due to the presence of the liquid: buoyancy and capillarity. For changes in liquid level on the order of 1 cm, the change in buoyancy force is approximately $100 \,  \mu$N for Fluorinert. This is slightly larger than the capillary force from the surface tension present in the experiment, which the Young-Dupr\'{e} equation specifies as $40 \, \mu$N for a 1 mm diameter thread and the material properties given in Table.~\ref{tab:liquid}.

\subsection{Force balance equations from virtual work \label{subsec:virtual}}

As shown in Fig.~\ref{fig:exp}, the partially immersed thread in the final state is held  up by an external force $2 \pi R {F}_{\mathrm{ext}}$, where  $F$ has units of surface tension (mN/m). Physically, this external force is the sum of both  capillary and  buoyancy effects, the weight of the partially immersed thread, and the effective weight of the immersed metal weight.
We calculate $2 \pi R {F}_{\mathrm{ext}}$ by the principle of virtual work, determining the change in the potential energy as we lift the thread by an infinitesimal distance  $\delta$ above the current state:
\begin{equation}
\left. 2 \pi R \ {F}_{\mathrm{ext}}= \frac{\mathrm{d} \Delta U}{\mathrm{d} \delta} \right|_{\delta \rightarrow 0}
\label{equ:virtualwork}
\end{equation}
where $\Delta U$ is calculated from the sum of the surface and gravity potential energies: $\Delta U= \Delta U_{\mathrm{surface}} + \Delta U_{\mathrm{gravity}}$. 

For values of the surface energy at the solid-air ($\gamma_{sg}$) and solid-liquid ($\gamma_{sl}$) interfaces, the total surface energy changes by 
\begin{equation}
\Delta U_{\mathrm{surface}}(\delta) = 2 \pi R \left[({L}_{sg}+\delta)\gamma_{sg} + ({L}_{sl}-\delta)\gamma_{sl}\right]
\label{equ:U_surface}
\end{equation}
Differentiating Eq~\eqref{equ:U_surface}, the capillary force arising from the surface energy is
\begin{equation}
\left. \frac{\mathrm{d} \Delta U_{\mathrm{surface}}}{\mathrm{d} \delta}\right|_{\delta \rightarrow 0}= 2 \pi R (\gamma_{sg} - \gamma_{sl})= 2\pi R \gamma \cos\theta,
\label{equ:surface_force}
\end{equation}
with $\gamma$ (see Table \ref{tab:liquid}) the surface tension at liquid-air interface, and the last equality coming from the Young-Dupr\'e equation. 

The potential energy also changes due to an increase in gravitational potential energy as the thread raises by a distance $\delta$, determined from both the thread and the metal mass. This change is 
\begin{equation}
\Delta U_{\mathrm{gravity}} = \pi  R^{2} g \int_{0}^{\delta}\rho(z) h(z)\mathrm{d} {z}
+ m^*g \int_{0}^{\delta}\mathrm{d}{z}
\label{equ:U_gra}
\end{equation}
The first term on the right side is an integral of the thread density over the whole length of the thread. Above the fluid surface, the thread is in the air, so the density is $\rho=\rho_s$. Below the fluid surface, the correct density for the calculation is the the effective density $\rho^* \equiv \rho_{s}-\rho_{l}$. This gives 
$$
\rho(z) h(z) = \left [ \rho_{s}({L}_{sg}+{z}) + \rho^*({L}_{sl}-{z}) \right]    
$$
The second term on the right side in Eq.~\eqref{equ:U_gra} is the fixed metal mass. Therefore, the gravitational force is 
\begin{equation}
\left. \frac{\mathrm{d} \Delta U_{\mathrm{gravity}}}{\mathrm{d} \delta}\right|_{\delta \rightarrow 0}
= \pi R^{2} g (\rho_{s}L-\rho_{l}{L}_{sl})+m^*g.
\label{equ:gravity}
\end{equation}

Adding Eq.~\eqref{equ:surface_force} and \eqref{equ:gravity} to form the right side of Eq.~\eqref{equ:virtualwork}, we have an expression for the external force holding up the thread:
\begin{equation}
2 \pi R \ {F}_{\mathrm{ext}} = 2\pi R \gamma \cos \theta + \pi R^{2} g (\rho_{s}L-\rho_{l}{L}_{sl})+m^*g.
\label{equ:ext}
\end{equation}

For a given cross section of the thread immersed in the liquid, we assume the cross section is subject to a force $2 \pi R {F}_{\mathrm{imm}}$,\cite{cite:Marchand} which equals the sum of gravity, buoyancy, and surface stress at the solid-liquid interface. Similar to Eq.~\eqref{equ:ext}, one has
\begin{equation}
-2\pi R \ {F}_{\mathrm{imm}}= 2 \pi R \Gamma - \pi R^{2} g (\rho_{s}L-\rho_{l}{L}'_{sl}) - m^*g
\label{equ:imm}
\end{equation}
where the first term on the right side contains an unknown parameter $\Gamma$ that comes from the change of the thread surface from air to liquid. Unlike Eq.~\eqref{equ:ext}, which comes from principle of virtual work, Eq.~\eqref{equ:imm} comes from the balance of stress. Therefore, instead of surface energy, $\Gamma$ is the change of the surface stress from solid-air to solid-liquid interface. The second and the third term in Eq.~\eqref{equ:imm} come from gravity and buoyancy on both the thread and the metal weight. Note that this assumption leaves open the possibility of a tangential component of the contact line force given by
\begin{equation}
2\pi R \ F_{\mathrm{\parallel}} = 2\pi R \left({F}_{\mathrm{ext}}-{F}_{\mathrm{imm}}\right)= 2\pi R\left(\gamma \cos\theta + \Gamma\right)
\label{equ:tan}
\end{equation}
This expression allows us to measure the value of $\Gamma$, and thereby determine whether or not a non-zero $F_{\parallel}$ is present: if $\Gamma= -\gamma\cos\theta$, then $F_{\parallel}=0$ and the surface stress of the gel behaves as if it were the surface tension of a liquid. 

\subsection{Displacement fields}

To determine $\Gamma$, we follow the method suggested by \citet{cite:Marchand}: we first measure the displacements of the beads, and then fit the observed displacement data to a theoretical prediction for  $\Gamma$. 
In \S\ref{subsec:virtual}, we derived the force balance equation for a partially immersed thread. Now we compare the force difference before and after the liquid level increases, which corresponds to the initial state (the left thread in Fig.~\ref{fig:exp}) and the final state (right thread). We start with the stress tensors and the strain fields in two states. Then we derive the displacement field by integrating the strain difference between the two states.

From \S\ref{subsec:virtual}, we know all of the applied forces on the thread, which we can use to calculate the stress tensor inside the thread. This  includes contributions from not only the forces in the vertical direction ($z$ or $z'$), but also radial stresses which arise from the surface stress. Again, all primed variables refer to the initial state, and all un-primed variables to the final state.

We begin with the stress tensor in the final state: above the liquid surface (${z}>0$),  $ {\sigma}_{rr}=0 $ as 
there is no hydrostatic pressure or the change in surface interface. Below the liquid surface (${z}<0$), we assume that the hoop stress is 
\[
{\sigma}_{rr}= -\frac{\Gamma}{R} + \rho_l g {z} 
\]
where the first term on the right side comes from the change in surface stress from air to liquid, and the second term comes from hydrostatic pressure.  Dividing Eq.~\eqref{equ:ext} and Eq.~\eqref{equ:imm} by the cross section of the thread ($\pi R^2$), we describe both the vertical and radial stresses applied to the thread:
\begin{equation}
{\sigma}_{zz}({z})= 
\begin{cases}
\frac{\displaystyle 2\gamma\cos\theta}{\displaystyle R} - g \rho_l {L}_{sl}+\alpha, & {z}>0 \\
-\frac{\displaystyle 2\Gamma}{\displaystyle R} - g \rho_l {L}_{sl}+\alpha, & {z}<0 
\end{cases} 
\label{equ:stress_z}
\end{equation}
where $\alpha\equiv g \rho_s L + m^{*}g/\pi R^2$ is a constant, and
\begin{equation}
{\sigma}_{rr}({z})= 
\begin{cases}
0, & {z}>0 \\
-\frac{\displaystyle \Gamma}{\displaystyle R} - \rho_l g {z}, & {z}<0
\end{cases}
\label{equ:stress_r}
\end{equation}
The stress tensor for the initial state takes the same form, but with $z$ and $L$ having primes.

Assuming a linearly elastic, incompressible material with $L\gg R$ (conditions reasonable for our thread), the vertical displacements $u_z$ arise solely from $\epsilon_{zz}$:
\begin{equation}
\partial_z u_z= \epsilon_{zz}= \frac{1}{E} \left(\sigma_{zz}-\sigma_{rr}  \right)
\label{equ:strain}
\end{equation}
A detailed derivation of the strain fields for both states is provided in appendix \S \ref{subsec:strainderive}. Since our experimental measurements arise from comparing the displacements for two different states, we want to model the difference of the strain fields, $\Delta \epsilon= \epsilon_{zz} - {\epsilon}'_{zz}$. We use the transformation ${z}'=z+\Delta L$ and ${L}'_{sl}= L_{sl} - \Delta L$ in the calculations that follow. The strain difference in Segment A (${z}'>\Delta L$, $z>0$, always above the liquid) is
\begin{equation}
\Delta \epsilon_A= \frac{1}{E} \left[ -\rho_l g ( {L}_{sl} - {L}'_{sl} ) \right] = -\frac{\rho_l g \Delta L}{E}.
\label{equ:strainA}
\end{equation}
Note, this value is a constant, and all of the values are known/measured from separate experiments (no free parameters). The strain difference in Segment B (${z}'>0$, $z<0$, switches from above to below the liquid) is
\begin{equation}
\Delta \epsilon_B
=\frac{1}{E} \left[ -\frac{2\gamma\cos\theta+\Gamma}{R} - \rho_l g(z + \Delta L) \right].
\label{equ:strainB}
\end{equation}
Note that this predicts a position-dependent change in strain, since the hydrostatic pressure increases along the length of the immersed thread.

To determine the displacement field, we integrate Eq.~\eqref{equ:strainA} and \eqref{equ:strainB} along the vertical direction and obtain
\begin{equation}
\Scale[0.8]{
u_z\left(z\right)= 
\begin{cases}
\frac{\displaystyle \rho_l g \Delta L}{\displaystyle E} z, & z>0 \\ 
-\frac{\displaystyle 2\gamma\cos\theta+\Gamma}{\displaystyle ER}z - \frac{\displaystyle \rho_l g \Delta L}{\displaystyle E} z - \frac{\displaystyle \rho_l g}{\displaystyle 2E}z^2, & -\Delta L<z<0
\end{cases} } 
\label{equ:dis}
\end{equation}
where the displacement field is referenced to the position in the final state. Notice that displacements are always compressive in the vertical direction: $u_z$ is negative in Segment A since $z$ is positive, while $u_z$ is positive in Segment B since $z$ is negative.
Note that within Segment A, the displacements of the beads arise due only to buoyancy, for which there are no free parameters. This allows us to use data from this region to measure the magnitude of the swelling effects in the experiments. 
However, in Segment B, the displacement of the beads arises from two effects: capillary forces (both surface tension and surface stress) and buoyancy. 
Therefore, we can determine the only unknown quantity, the change in surface stress $\Gamma$, by measuring  buoyancy, the surface tension $\gamma \cos\theta$, and the displacements of beads $u$. By fitting this data to Eq.~\eqref{equ:dis}, Eq.~\eqref{equ:tan} allows us to  calculate $F_{\mathrm{\parallel}}$, where $\gamma \cos\theta$ is known, and $\Gamma$ is obtained from fitting.

\section{Results \label{sec:result}}

\subsection{Equilibration time \label{sec:time}}

\begin{figure}
\begin{center}
\includegraphics[width=0.45\textwidth]{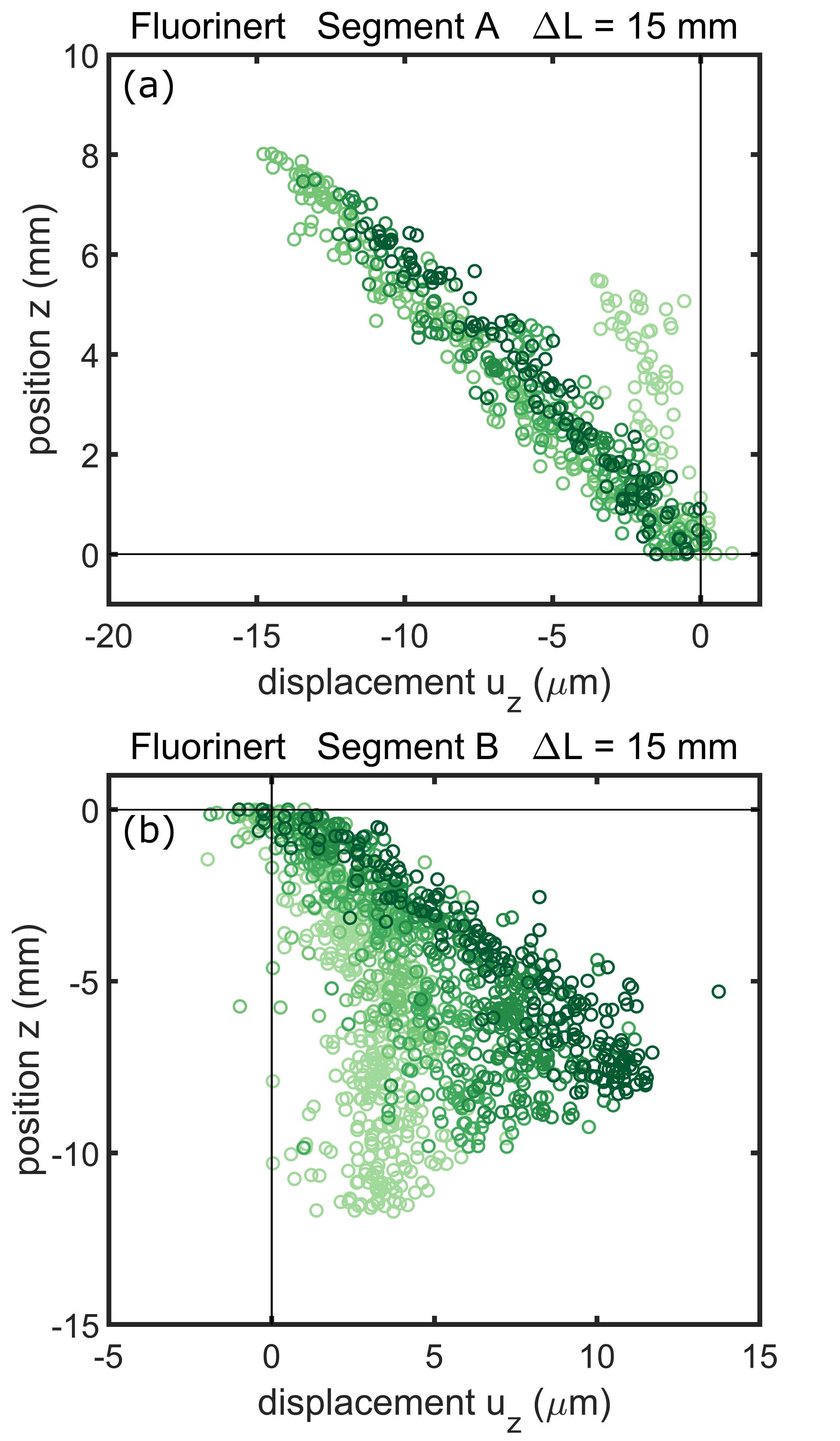}
\end{center}
\caption{Dependence of the measured displacement field $u_z$ on the waiting time $t_w=  0$ (light green) $6, 12, 18, 24$~hrs (dark green). Data is shown (a) in Segment A, which is always above the liquid surface and (b) in Segment B, which switches from above to below the liquid surface. The negative slope indicates a net compressive stress on the immersed thread.}
\label{fig:time}
\end{figure}

To test the effect of waiting time on the deformation field, we perform experiments using Fluorinert, which has been reported to induce little swelling\cite{cite:Whitesides}. We repeat an identical experiment using a liquid level change of $\Delta L = 15$~mm, for five different waiting times $t_w$, up to one day in duration. As shown in Fig.~\ref{fig:time}a, the displacements of the beads in Segment A for all $t_w > 6$~hrs are statistically identical. Therefore, beads which remain above the liquid surface reach equilibration within 6 hrs. Within Segment B (see Fig.~\ref{fig:time}b), the displacements of the beads continue to evolve up to $t_w = 18$~hrs, after which the displacements of beads below $z=-5$~mm are consistent with the values observed at 24 hrs. Therefore, the lower portion of Segment B reaches equilibration within about 1 day. 

This long equilibration time is consistent with previous experiments done on a droplet deposited on a thin film\cite{cite:Alain_1996, cite:Park_visual, cite:Kajiya, cite:Karpitschka_move}. In these experiments, the surface of the thin film deforms at a rate of several microns per minute. Since our thread is several centimeters long, we expect timescales about a thousand times longer (hours). In contrast, \citet{cite:Park_visual} showed that the contact angle of a droplet on a thin film develops within less than a second, via a rapid elastic deformation of the substrate. 
This sets a clear separation of timescales for our experiment: with the contact angle between the thread and the liquid being determined within a second by elastic deformation, while the internal strain field equilibrates over many hours by viscous deformation. 
To further test these viscoelastic traits, we cast a block of PVS and measure its complex shear modulus using a rheometer (Anton Paar MCR-302). We find the loss modulus to be  ${G}'' \approx 7$~kPa, while the storage modulus is ${G}'\approx 30$~kPa. As such, we expect to observe a slow, viscous motion as well as an elastic response.

\subsection{Swelling \label{sec:swelling} }

\begin{figure}
\begin{center}
\includegraphics[width=0.45\textwidth]{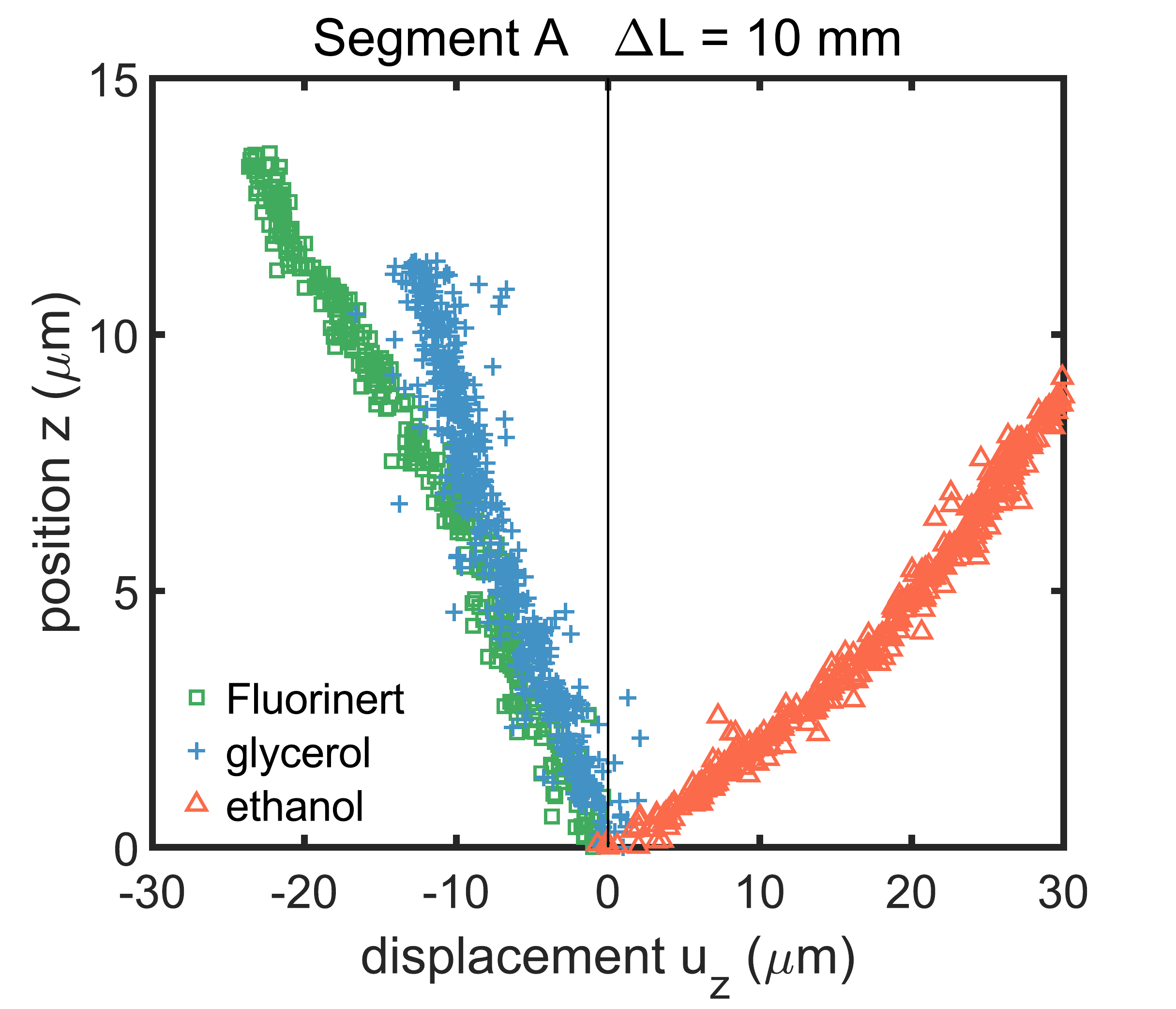}
\end{center}
\caption{Measured displacement $u_z$ for the three liquids considered, all taken within Segment A (above the liquid surface), where the results are less sensitive to waiting time. For glycerol and Fluorinert, $t_w= 12$~hrs; for  ethanol $t_w= 6$~hrs (to minimize the influence from evaporation).}
\label{fig:swelling}
\end{figure}

Because our model (\S\ref{sec:theo}) doesn't account for swelling effects, we need to both determine under which conditions they are present, and adapt the model to account for their presence.  
Because all of the parameters in Eq.~\eqref{equ:dis} are known for $z>0$, we are able to directly compare the displacements of beads in Segment A with the theoretical prediction. Any difference between the data and the theory will identify the magnitude of swelling effects. As an additional benefit, recall that the bead displacements in Segment A are less sensitive to waiting time (see Fig.~\ref{fig:time}).

We perform experiments to measure bead displacements for threads immersed each of the three liquids, for a liquid level change of $\Delta L = 10$~mm; these results are shown in Fig.~\ref{fig:swelling}. As given by Eq.~\eqref{equ:dis}, the buoyancy force compressing the thread increases as we increase the liquid level; we therefore expect $u_z <0$ for beads at locations $z>0$. We observe this compressive effect for experiments done in both glycerol and Fluorinert. However, for experiments done in ethanol, we observe $u_z >0$ and attribute this to swelling (as would be expected from the work of \citet{cite:Whitesides}) Although Segment A is not immersed in liquid, the ethanol absorbed within Segment B  can also diffuse into Segment A. This causes the whole thread to swell enough to overcome the compression due to buoyancy, and appear as if it is in tension ($u_z > 0$).

\begin{figure}
\begin{center}
\includegraphics[width=0.45\textwidth]{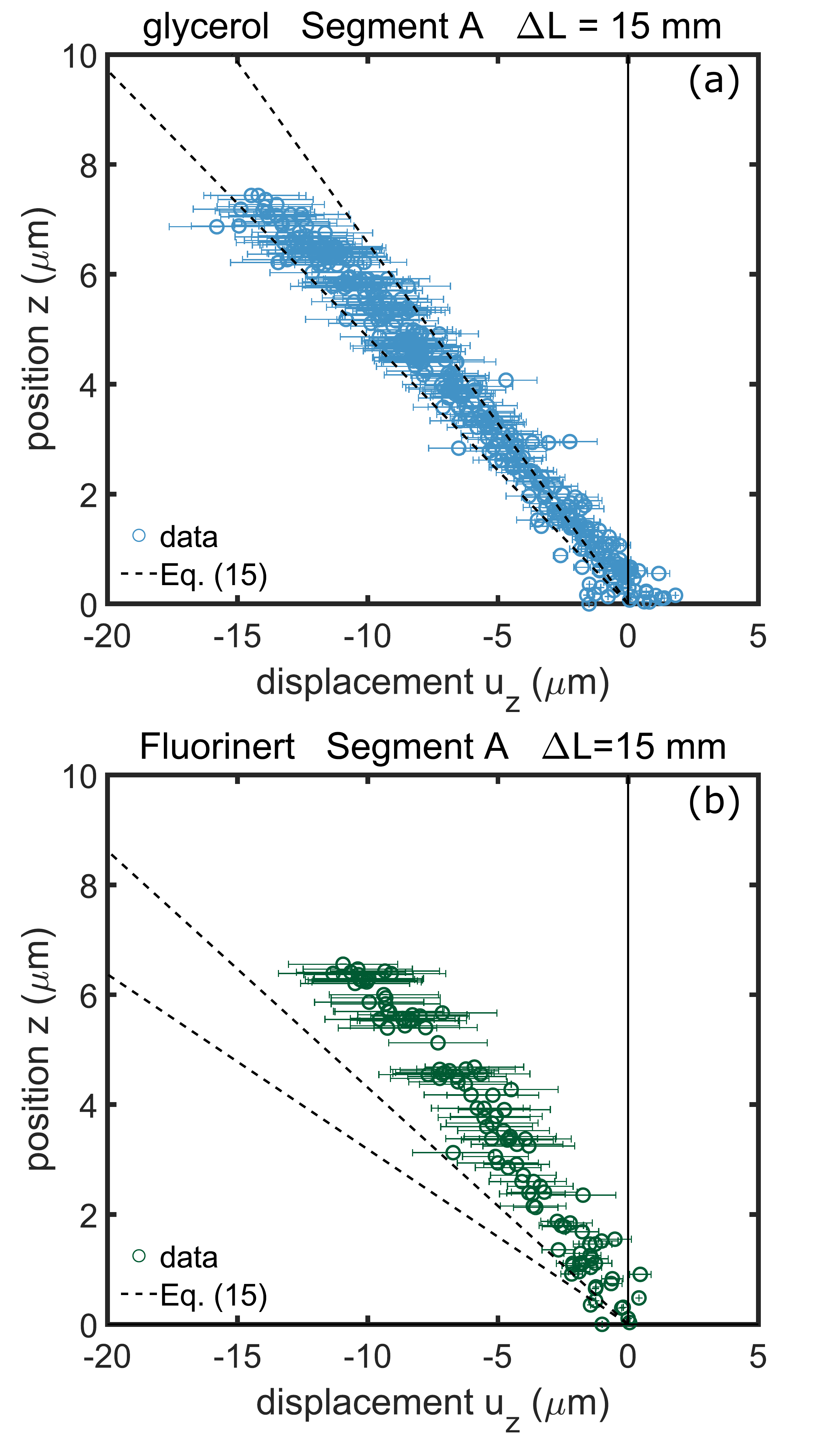}
\end{center}
\caption{Measured displacement $u_z$ in Segment A (above the liquid surface), compared to predicted values from Eq.~\eqref{equ:dis} where the pair of dashed lines are the upper and lower bounds of the theoretical prediction, calculated from the standard error on the Young's modulus measurement. (a) Results for glycerol, $t_w= 12$~hrs. (b) Results for Fluorinert, $t_w = 24$ hrs.}
\label{fig:swell2}
\end{figure}

Using experiments done in glycerol and Fluorinert, we can perform a direct comparison to the model using the known parameters provided in Table~\ref{tab:liquid}. In Fig.~\ref{fig:swell2}, we observe that data obtained in glycerol is in agreement with the model, while for Fluorinert the $u_z$ displacements are slightly less than predicted. Data for $u_r$ displacements (for experiments in glycerol) are shown in the Appendix (Fig.~\ref{fig:radial} in \S\ref{subsec:radial}) and are also in agreement with predictions. From this, we conclude that there are no swelling effects due to the gylcerol, that buoyancy (alone) controls the changes to strain within Segment A, and that the assumption of constant total length $L$ and radius $R$ are valid. 

For experiments done in Fluorinert, we need to adapt the model to account for the small swelling effect. We directly measure the swelling of PVS immersed in Fluorinert, in separate experiments on a centimeter-sized PVS block. As reported in detail in the Appendix (\S\ref{subsec:ratio}), we observe a weight change corresponding to a volume-swelling ratio  $\approx 0.2 \%$ (and none for glycerol), which is consistent with the discrepancy shown in Fig.~\ref{fig:swell2}b. Because the swelling occurs within 15 minutes, this timescale is well-separated from the  hour-long equilibration time for the bead position measurements. 

Note that although we use PVS for our experiments, the swelling effect we observe is consistent with previous elastocapillary studies using polydimethylsiloxane (PDMS) substrates. Under room temperature, \citet{cite:Whitesides} found that the volume of PDMS immersed in ethanol increased $4\%$, but by less than  $1\%$ for glycerol, water, ethylene glycol, or perfluorocarbon liquids (such as Fluorinert). Importantly, since the swelling ratio changes as a function of temperature,\cite{cite:style_swell} it is important to fix the experimental conditions in such experiments.

\subsection{Electrostatic forces}

In our model, we excluded electrostatic forces, which can arise due to triboelectric charging when the thread is removed from the silanized capillary tube. Such forces are observed in our experiments before the intial-state measurements, when the thread is electrically-attracted to the wall of the glass cuvette. Polar solvents such as glycerol could shield these forces, while nonpolar liquids like Fluorinert have limited shielding ability\cite{book:Russel}. To determine  whether electrostatic forces are present in our experiments, we neutralize the charges on the surface of the thread and the cuvette with an anti-static gun (Zerostat 3, Milty\texttrademark), and observe the same displacement results as presented in Fig.~\ref{fig:swell2}b. Therefore, we determine that there is no significant influence from electrostatic charges on the results of our experiments.

\subsection{Tangential component of the contact line force \label{sec:tangent}}

Finally, we are able to use the results our experiments on waiting time, swelling, and electrostatics to determine whether there is a measurable tangential component to the contact line force $F_{\mathrm{\parallel}}$ in Eq.~\eqref{equ:tan}. To select a liquid, we consider several requirements. To avoid both significant swelling and a high evaporation rate, we need to select either glycerol (no swelling) or Fluorinert (small swelling). However, glycerol has an additional difficulty: its lack of optical clarity makes it unsuitable for measurements in Segment B, where images must be collected below the liquid surface. This leads us to a choice of Fluorinert, which is optically clear, with the caveat that we will then need to account for the measured swelling ratio by adapting the model.

Since the swelling ratio is small, we approximate its effects in Eq.~\eqref{equ:dis} as being constant throughout the whole thread. We modify this equation to include a swelling parameter $\beta$ which takes the same value in both Segment A and Segment B. 
\begin{equation}
\Scale[0.8]{
\begin{aligned}
u_z + u_{A} - \beta z &= - \frac{\rho_l g \Delta L}{E}z, & z>0 \\ 
u_z + u_{B} - \beta z &= -\frac{2\gamma\cos\theta+\Gamma}{ER}z - \frac{\rho_l g \Delta L}{E} z  - \frac{\rho_l g}{2E} z^2, & z<0
\end{aligned}}
\label{equ:dis_beta}
\end{equation}
We have additionally included a constant offset to each displacement field ($u_A$ and $u_b$) so account for measurement uncertainty in the determination of the location of the  liquid surface ($z=0$), due to the presence of the miniscus. 
The unknown parameters in Eq.~\eqref{equ:dis_beta} are the swelling parameter $\beta$ and a constant material property $\Gamma$ which represents the change in surface stress from immersion in air vs. liquid. 

This gives us two equations, and two fitting parameters, for each experimental run. We determine the optimal values for $\beta$ and $\Gamma$ by simultaneously fitting (MATLAB  function \texttt{lsqnonlin}) the data from Segments A and B, for two separate experiments.  The two experiments were done  with  liquid level changes  $\Delta L= 15$~mm and $\Delta L= 20$~mm. Because the difference in the liquid level between these two experiments (5 mm) corresponds to 20\% of the total thread length $L$, the amount of fluid in contact with the thread is quite different and the magnitude of the swelling is  not necessarily the same. Therefore, we allow $\beta$ to be different for the two experiments, but we expect $\Gamma$ to match.

\begin{figure}
\begin{center}
\includegraphics[width=0.45\textwidth]{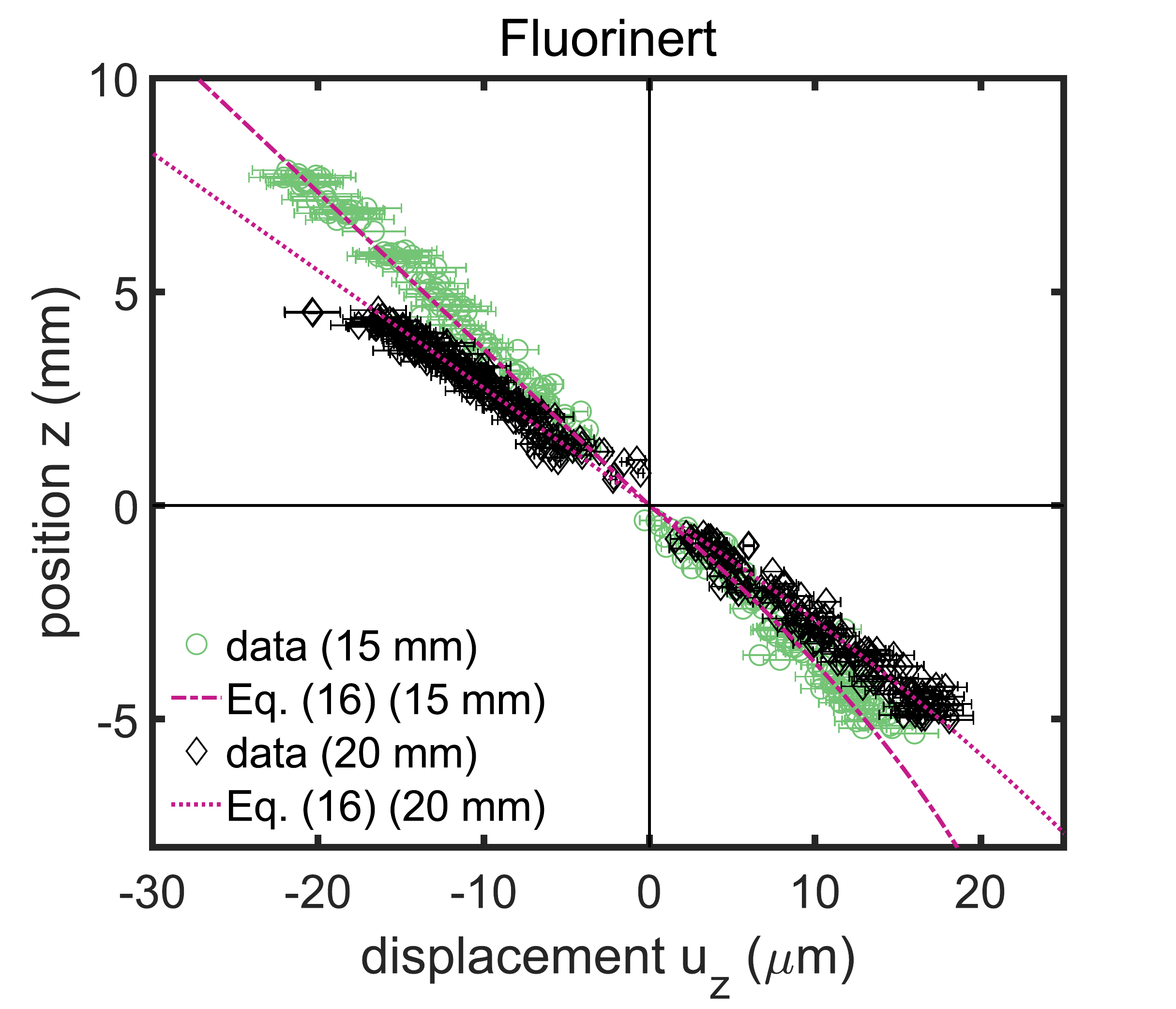}
\end{center}
\caption{Measured displacement $u_z$ in Segment A (above the liquid surface) and Segment B (below the liquid surface), compared to predicted values from Eq.~\eqref{equ:dis_beta}. The continuity of the fit across the two equations is guaranteed by the simultaneous fitting process. The two datasets are for $\Delta L = 15$~mm  (green circles) and $20$~mm (black diamonds), both with a waiting time $t_w = 24$~hrs. Fitting parameters are discussed in the text.}
\label{fig:tangent}
\end{figure}

The data and fitting results are shown in Fig.~\ref{fig:tangent}. We find that the best-fit swelling ratio is $\beta= 0.11 \%$ for $\Delta L = 15$~mm and $\beta= 0.24 \%$ for $\Delta L = 20$~mm. These values are both consistent with our direct measurements of the swelling ratio (see \S\ref{sec:swelling} and Supp. Matt.) We determine that the surface stress consistent with both sets of measurements is $\Gamma= -12.8 \pm 10.0$~mN/m ($95\%$ confidence interval).
Using the known values of $\gamma$ and $\theta$ in Eq.~\eqref{equ:tan}, this gives the measured value of the tangential component of the contact line force
$$
F_{\parallel} = 1.2 \pm 10 \ \mathrm{mN/m},
$$ 
which is consistent with $F_{\parallel} =0$. As a comparison, the reported value in previous liquid-thread experiments\cite{cite:Marchand} (using ethanol as the immersion liquid) is $F^\mathrm{eth}_{\parallel} = 43.3 \pm 8.2$~mN/m. Using values from Table~\ref{tab:liquid}, we normalize $F_{\parallel}$ and $F^\mathrm{eth}_{\parallel}$ by the surface tension $\gamma$ for each liquid. The results are  $F_{\parallel}/\gamma_{\mathrm{Fluorinert}} = 0.1 \pm 0.6$ and $F^\mathrm{eth}_{\parallel}/\gamma_{\mathrm{ethanol}} = 2.0 \pm 0.4$. Because these two normalized values are mutually outside  each other's confidence intervals, we conclude that our value of $F_{\parallel}$ disagrees with what was measured previously.

Several explanations are possible for the discrepancy between the measurements of the contact line forces in the two cases.  First, ethanol causes swelling during the observation period (see \S\ref{sec:swelling}). Second, if data were recorded quickly (to minimize this effect, and also the effect of evaporation), the beads would not yet have reached their equilibrium positions (see \S\ref{sec:time}). Third, in both cases the models are based on linear elastic theory. However, when the gel thread experiences swelling,  gel porosity should also be taken into account\cite{cite:Daniels, cite:LeeJB, cite:Reis, cite:Kajiya_2011}.
Finally, the measured strain in our experiments is an order of magnitude smaller than was present in the \citet{cite:Marchand} experiments. It is unknown what effect the magnitude of strain has on the magnitude of the tangential component of the contact line force. 

Recent papers \cite{cite:Xu_shuttle, cite:Andreotti_shuttle, cite:Snoeojer_shuttle, cite:Style_stress, cite:Xu_surface, cite:Schulmann_shuttle} aim to disentangle how the surface stress of an elastomer changes under tangential stresses. This effect, known as the Shuttleworth effect \cite{cite:Shuttleworth} argues that the solid surface stress changes in response to any applied tangential stress.  If the strains of the substrate are different for  solid-air vs. solid-liquid surfaces, the Young's angle $\theta$ might change due to the Shuttleworth effect. It has been shown that the Young's angle for PVS gel does not change when a uniform strain is applied to the substrate\cite{cite:Schulmann_shuttle}: the vertical strain due to our hanging weight $m^*$ does not affect Young's angle. Second, the strain of the thread in Segment A  comes only from  buoyancy, while in Segment B, the strain comes from both capillarity and  buoyancy. Therefore, there is a strain difference between the solid-air surface and the solid-liquid surface. Our results suggest that $\Gamma \approx -\gamma\cos\theta=-\gamma_{sg}+\gamma_{sl}$, such that the Young's angle $\theta$ remains constant. This is consistent with the expected lack of a Shuttleworth effect in our system.

\section{Conclusions}

We perform controlled experiments to quantify the elastocapillary effects on a PVS gel thread partially immersed in a variety of liquids. We show that swelling and buoyancy have effects of a similar magnitude to capillary forces.  We also demonstrate that the differing swelling ratios among various liquids render some as poor choices for elastocapillary experiments: glycerol and Fluorinert are favored over ethanol. In the worst cases,  swelling can overwhelm what would have otherwise been compressive forces.  However, glycercol also presents difficulties due to a lack of optical clarity. Because Fluorinert has a swelling ratio of less than $1\%$, as well as excellent optical clarity and slow evaporation, we find it to be a good choice for elastocapillary experiments.

\begin{figure}
\begin{center}
\includegraphics[width=0.45\textwidth]{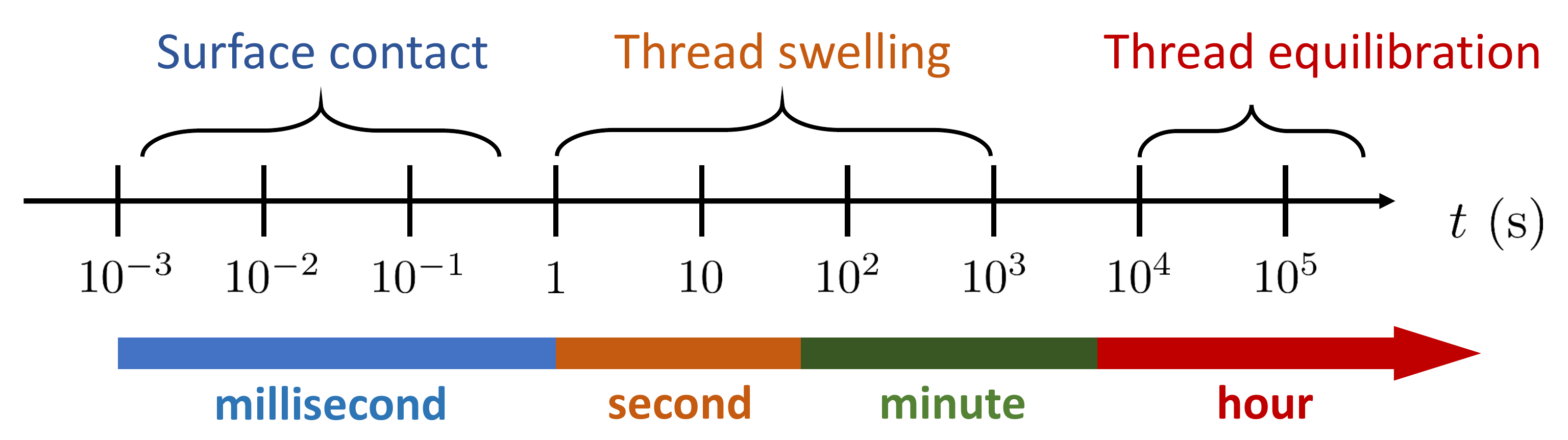}
\end{center}
\caption{Comparison of the timescales for various dynamical processes observed in elastocapillary experiments. The surface contact timescale is from \citet{cite:Park_visual}; other values are measured here.}
\label{fig:timescale}
\end{figure}

In quantifying each of these effects, we observed several distinct timescales, shown in Fig.~\ref{fig:timescale},  that need to be considered when undertaking experimental designs.  In our experiments, the beads in the thread take hours to reach their equilibrium positions, while the deformations due to capillarity and swelling take less than an hour. 

Finally, guided by these choices of material and timescale, we performed experiments in which only two free parameters remained: a swelling ratio $\beta$ and the change in surface stress $\Gamma$.  Our results show that the tangential component of the contact line force is likely zero, as suggested by the results of droplet-film experiments\cite{cite:Style_angle,cite:Bostwick,cite:Style_2012,cite:Park_visual}. Moreover, we observe that the solid surface stress remains unchanged under tangential compression, suggest that a Shuttleworth effect is not present for this system, consistent with prior work\cite{cite:Schulmann_shuttle, cite:liang_2018, cite:Masurel_2018}.

\section*{Acknowledgements}

We are grateful for financial support from the  National Science Foundation under Grants DMR-1608097 (experiments, SYC and KED) and DMS-1517291 (theory, AB and MS).

\bibliography{refer2}


\newpage
\section{Appendices}

\subsection{Strain derivation \label{subsec:strainderive}}

Due to the cylindrical geometry of the thread, we choose a cylindrical coordinate system. The linear stress tensor is given by
\begin{equation}
\sigma_{ij}=\frac{E}{1+\nu} \left[\epsilon_{ij} + \frac{\nu}{1-2\nu}\epsilon_{kk}\delta_{ij} \right]
\end{equation}
where $E$ is the elastic modulus, $\nu$ is the Poisson's ratio, $\epsilon_{ij}$ is the strain tensor with $i,j$ as $(r, \phi, z)$, $\epsilon_{kk}=\epsilon_{rr}+\epsilon_{\phi\phi}+\epsilon_{zz}$ is the trace, and $\delta_{ij}$ is the Kronecker delta function. Formulas for the strains are given by
\begin{equation}
\epsilon_{rr}=\partial_{r}u_{r}, \quad \epsilon_{\phi\phi}= \frac{u_r}{r}+\frac{1}{r}\partial_{\phi}u_{\phi}, \quad \epsilon_{zz}=\partial_{z}u_{z}
\end{equation}
where $u_i$ is the displacement of a given position in the thread. In the experiment, $u_i$ is the displacements of the beads. We assume the solution is axisymmetric, as no torque is applied to the thread. Thus, all variations with respect to angle $\phi$ can be set to zero. For a small radius thread ($R \ll L$), $\epsilon_{rr} \approx \epsilon_{\phi\phi}$ and
the trace of the strain tensor $\epsilon_{kk}$ can be written as
\begin{equation}
\epsilon_{kk} \approx 2\epsilon_{rr} + \epsilon_{zz}
\label{equ:double}
\end{equation}

Hence, $\epsilon_{rr}=\frac{1}{2}(\epsilon_{kk}-\epsilon_{zz})$. The following calculation then isolates an expression for axial strain $\epsilon_{zz}= \partial_z u_z$:
\begin{equation}
\Scale[0.8]{
\begin{aligned}
\sigma_{zz}-2\nu\sigma_{rr} 
&= \frac{E}{1+\nu} \left[ \left( \partial_z u_z - 2\nu\partial_r u_r \right) + \frac{\nu}{1-2\nu}\left(1-2\nu\right)\epsilon_{kk} \right] \\
&= \frac{E}{1+\nu} \left[ \left( \partial_z u_z - \nu\left(\epsilon_{zz}-\partial_z u_z\right) \right) + \nu \epsilon_{kk} \right]\\
&= E \partial_z u_z
\end{aligned}}
\end{equation}
We take our thread to be incompressible ($\nu = 1/2$) as assumed for most elastomer experiments, leading to Eq.~\eqref{equ:strain} in the main text.

By Eq.~\eqref{equ:stress_z}, \eqref{equ:stress_r}, and \eqref{equ:strain}, the strain field in the final state is
\begin{equation}
{\epsilon}_{zz}(r, {z})= 
\begin{cases}
\frac{\displaystyle 2\gamma\cos\theta}{\displaystyle ER} - \frac{\displaystyle \rho_l g {L}_{sl}}{\displaystyle E}, & {z}>0 \\ 
-\frac{\displaystyle \Gamma}{\displaystyle ER} - \frac{\displaystyle \rho_l g ({z}+{L}_{sl})}{\displaystyle E}, & {z}<0
\end{cases}
\label{equ:strain_prelim}
\end{equation}
Transforming ${L}_{sl} \to {L}'_{sl}$ and $z \to {z}'$, the strain field for the initial state is
\begin{equation}
{\epsilon}'_{zz}(r, {z}')= 
\begin{cases}
\frac{\displaystyle 2\gamma\cos\theta}{\displaystyle ER} - \frac{\displaystyle \rho_l g {L}'_{ls}}{\displaystyle E}, & {z}'>0 \\ 
-\frac{\displaystyle \Gamma}{\displaystyle ER} - \frac{\displaystyle \rho_l g \left({z}'+{L}'_{ls}\right)}{\displaystyle E}, & {z}'<0
\end{cases}
\label{equ:strain_prelim2}
\end{equation}

From incompressibility ($\nu=1/2$), Eq.~\eqref{equ:double} can be written as $\epsilon_{zz}+2\epsilon_{rr}=0$, giving a strain difference in radial direction as $\Delta \epsilon_{rr}= \Delta \epsilon_{zz} /2$. With Eq.~\eqref{equ:strainA} and \eqref{equ:strainB}, the displacement in the radial direction is 
\begin{equation}
\Scale[0.8]{
u_r\left(r, z\right)= 
\begin{cases}
\frac{\displaystyle \rho_l g \Delta L}{\displaystyle 2E} r, & z>0 \\ 
\left( \frac{\displaystyle 2\gamma\cos\theta+\Gamma}{\displaystyle 2ER} + \frac{\displaystyle \rho_l g (z+\Delta L)}{\displaystyle 2E}\right) r, & -\Delta L<z<0
\end{cases}}
\label{equ:rdis}
\end{equation}

\subsection{Displacement field in radial direction
\label{subsec:radial}}

\begin{figure}[h]
\begin{center}
\includegraphics[width=0.3\textwidth]{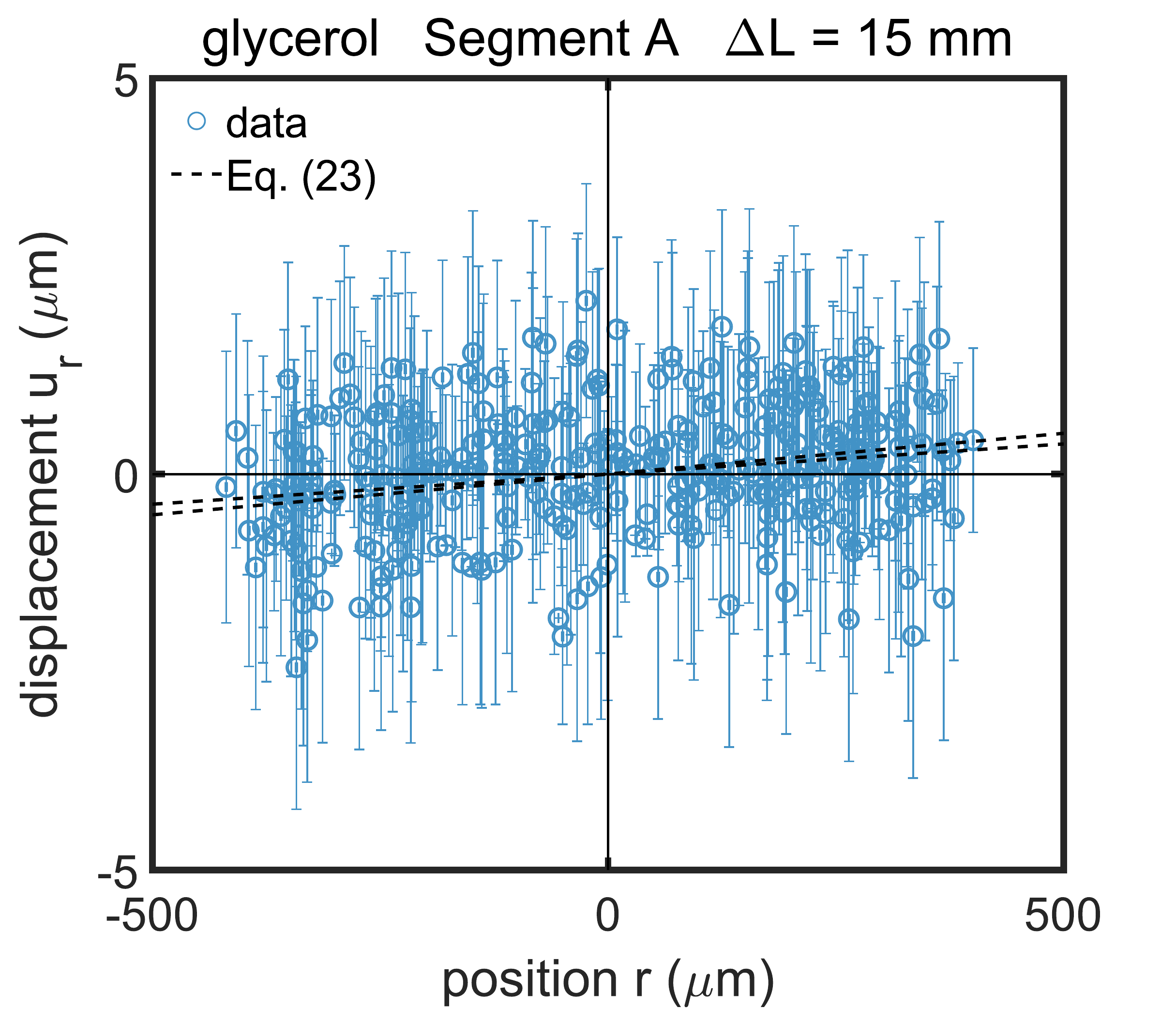}
\end{center}
\caption{Measured displacement $u_z$ in Segment A (above the liquid surface), compared to predicted values from Eq.~\eqref{equ:rdis} for glycerol, where the pair of dashed lines are the upper and lower bounds of the theoretical prediction, calculated from the standard error on the Young's modulus measurement. $t_w= 12$~hrs. }
\label{fig:radial}
\end{figure}

In Fig.~\ref{fig:radial}, we show that the displacement field $u_r$ is in agreement with the theoretical  prediction from Eq.~\eqref{equ:rdis}.  

\subsection{Measuring swelling ratio \label{subsec:ratio}}

\begin{figure}[h]
\begin{center}
\includegraphics[width=0.3\textwidth]{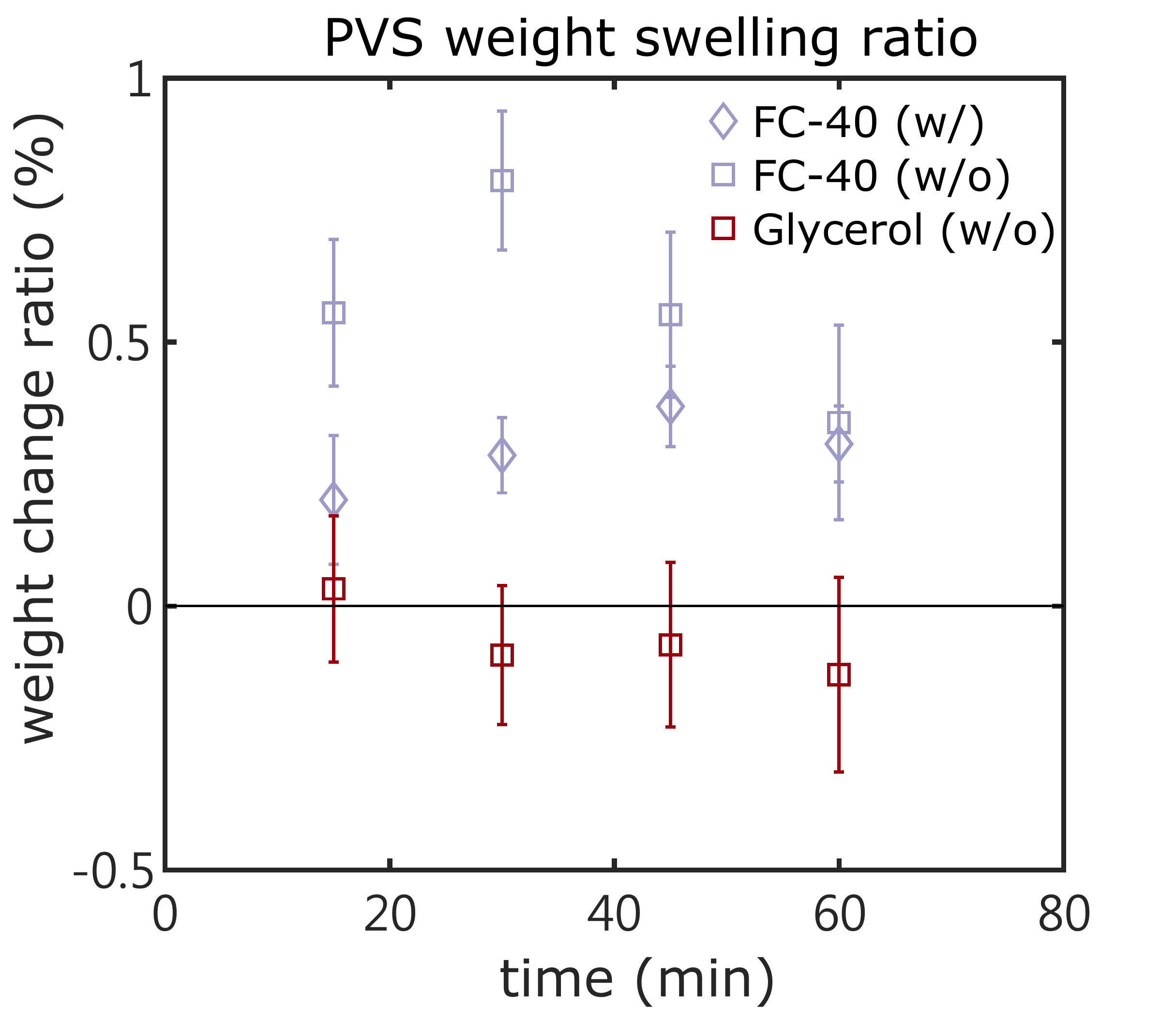}
\end{center}
\caption{PVS gel weight swelling ratio with different liquids. w/: gel with beads. w/o: gel without beads.}
\label{fig:ratio}
\end{figure}

To determine the swelling ratio for PVS immersed in our three liquids, we perform soaking experiments of various durations. Each experiment uses a PVS block, cast either with or without fluorescent beads, and cleaned as presented in \S\ref{subsec:clean}. Each block was weighed before and after  immersion for a duration up to 60 minutes, placed in either glycerol or Fluorinert FC-40. The blocks are dry initially, and are dried with an compressed air afterwards. The \%-difference in weight gives the swelling ratio.
The resulting measurements are shown in Fig.~\ref{fig:ratio}. Glycerol is observed to be non-swelling (no statistically-significant change in weight), consistent with the thread experiments in Fig.~\ref{fig:swell2}a. For Fluorinert, the measured swelling ratio depends on whether beads are present (swelling ratio $0.3\%$) or absent (swelling ratio $0.5\%$). The swelling occurs on timesclaes less than 15 min. Assuming the weight change is dominated by the absorption of liquid, the density ratio for 
PVS ($\rho_s = 1.040$~g/cm$^3$) and Fluorinert ($\rho_l = 1.855$~g/cm$^3$) predicts a volume change ratio $0.17 \%$
for the gel with beads.

\end{document}